\documentclass[english,a4paper, aps,prb,citeautoscript,twocolumn,showpacs,superscriptaddress]{revtex4-1} 

\usepackage[english]{babel} 

\usepackage[T1]{fontenc}
\usepackage[utf8]{inputenc}

\usepackage{mathptmx}
\usepackage{xcolor}

\usepackage{textcomp}
\usepackage{gensymb}

\usepackage[tbtags]{amsmath}
\usepackage{graphicx}
\usepackage[paperwidth=210mm,paperheight=297mm,centering,hmargin=1.6cm,vmargin=2.2cm]{geometry}
\usepackage{bm}

\newcommand{\WSe}{WSe$_\text{2}$}

\newcommand{\unit}[1]
{~\text{#1}}
\newcommand{\units}[1]
{\text{#1}}

\newcommand{\gt}{g^{(2)}}

\begin{document}

\title{Deterministic strain-induced arrays of quantum emitters \\
in a two-dimensional semiconductor}
\author{Artur Branny}
\author{Santosh Kumar}
\author{Raphaël Proux}
\author{Brian D. Gerardot}
\affiliation{Institute of Photonics and Quantum Sciences, SUPA, Heriot-Watt University, Edinburgh EH14 4AS, UK}

\date{\today}

\begin{abstract}
An outstanding challenge in quantum photonics is scalability, which requires positioning of single quantum emitters in a deterministic fashion. Site positioning progress has been made in established platforms including defects in diamond\cite{Mei2005, McLellan16} and self-assembled quantum dots\cite{Jons13, Yang16}, albeit often with compromised coherence and optical quality. The emergence of single quantum emitters in layered transition metal dichalcogenide semiconductors\cite{Srivastava15,He15,Koperski15,Chakraborty15,Tonndorf15,Kumar15,Kern16,Kumar16,Palacios16} offers new opportunities to construct a scalable quantum architecture. Here, using nanoscale strain engineering, we deterministically achieve a two-dimensional lattice of quantum emitters in an atomically thin semiconductor. We create point-like strain perturbations in mono- and bi-layer \WSe{} which locally modify the band-gap, leading to efficient funnelling of excitons\cite{Feng12, Castellanos13local, Li15} towards isolated strain-tuned quantum emitters\cite{Kumar15, Kern16} that exhibit high-purity single photon emission. These arrays of non-classical light emitters open new vistas for two-dimensional semiconductors in cavity quantum electrodynamics\cite{Vahala03} and integrated on-chip quantum photonics\cite{Lodahl15}.
\end{abstract}
\maketitle

Nanoscale strain engineering of the electronic band structure to create quantum confinement has long been pursued in bulk semiconductors. In particular, the exploitation of local elastic strain to laterally confine carriers has been achieved in epitaxially grown heterostructures such as quantum wells\cite{Kash89,Ober99,Schulein09} or wires\cite{Gershoni90}.  Unfortunately, robust strain-induced quantum confinement of carriers suitably isolated from detrimental surface states in bulk systems has not been realized due to the combination of: i) the limited elastic strain possible before plastic deformation and ii) small vertical strain propagation distances. Optically active two-dimensional (2D) semiconductors with a high elastic strain limit\cite{Bertolazzi11} offer renewed opportunities for nanoscale strain engineering of three-dimensional quantum confinement, which we pursue here.    

\begin{figure*}[t]
   	\centering
	\includegraphics[trim={1.4cm 18.2cm 1.4cm 1.4cm},clip, width=\textwidth]{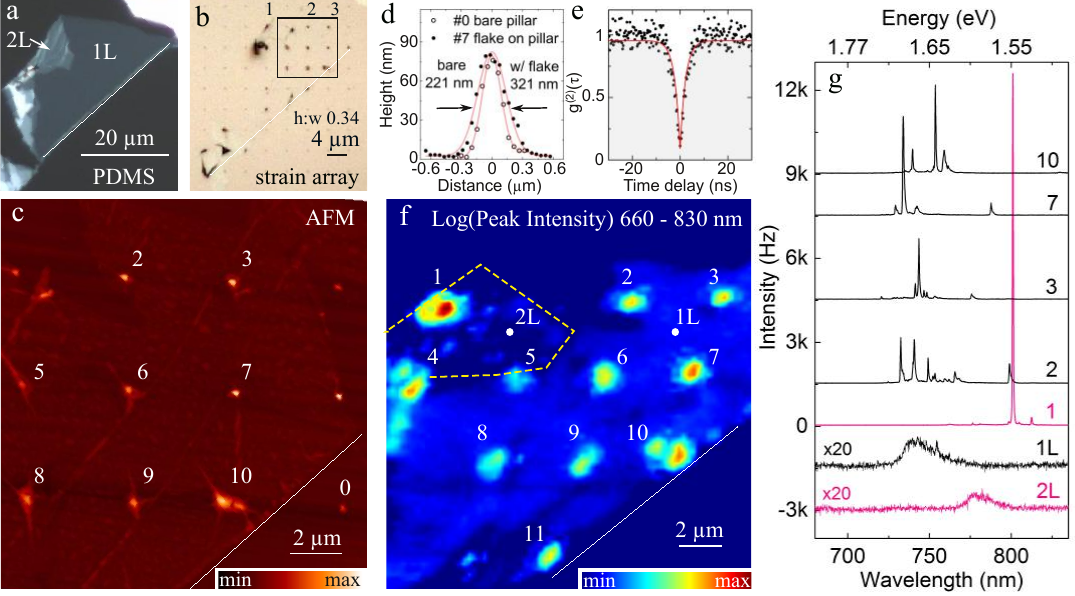}
   	\caption{Optical micrographs of an exfoliated 1L \WSe{} flake (a) before and (b) after transfer onto a Si substrate with an array of dielectric nanopillars. The black box in (b) identifies the nanopillars covered by the 1L region. (c) AFM image of the topography of the flake on top of the nanopillars, revealing a lattice of locally strained points. (d) Cross-section AFM profile of a bare nanopillar \#0 and nanopillar \#7 that is covered by the monolayer. (e) Second-order photon correlation statistics from a typical 1L quantum emitter revealing clear antibunching [$\gt(0) = 0.067 \pm 0.039$ and $\tau = 2.83 \pm 0.15 \unit{ns}$]. (f) Color-coded spatial map of the integrated PL signal in the spectral range of $660-830 \unit{nm}$. (g) Example PL spectra of isolated quantum emitters at the pillar locations as labelled. Also shown is the weak signal from the unstrained 1L and 2L \WSe{}.}
   	\label{fig:monolayer-small}
\end{figure*}

To achieve point-like strain perturbations in atomically thin \WSe{}, we use an all-dry transfer technique\cite{Castellanos-Gomez14} to transfer mechanically exfoliated flakes onto a substrate with a square lattice ($4\unit{\micro m}$ pitch) of dielectric nanopillars. This technique takes advantage of Van der Waals' forces to ensure the two-dimensional (2D) flake conforms to the contours of the patterned substrate and induces significant elastic strain at the locations of the nanopillars\cite{Reserbat14, Li15}. A \WSe{} flake consisting of mono-layer (1L) and bi-layer (2L) regions is shown in Figs.~\ref{fig:monolayer-small}a and \ref{fig:monolayer-small}b before and after transfer, respectively. In Fig.~\ref{fig:monolayer-small}b the nanopillars are identified by the change in contrast: the dark points in the micrograph correspond to locations of the nanopillars. The transferred flake’s topography is characterized by atomic force microscopy (AFM), as shown in Fig.~\ref{fig:monolayer-small}c-d. Scanning electron imaging of the same region (Supp. Info. Fig.~S1) confirms the same physical features. Figure~\ref{fig:monolayer-small}d compares the cross-section of a bare nanopillar (\#0 in Fig.~\ref{fig:monolayer-small}c) with an aspect ratio (height to width) of $\sim 0.3$ to that of a flake over a nanopillar (\#7 in Fig.~\ref{fig:monolayer-small}c). While the flake conforms quite closely to nanopillar \#7 without significant wrinkling, at most lattice positions the flake stretches over the nanopillar analogously to a canvas over a tent-pole. In particular, randomly oriented pleats emerging from nanopillars are observed as well as ripples which in some cases extend towards neighbouring nanopillars. These features are typical of 2D flakes suspended over corrugated substrates and can be further engineered\cite{Reserbat14}. Importantly, for the $4\unit{\micro m}$ pitch array used here, the wrinkles do not mask the point-like strain perturbation created by the nanopillars.

Hyperspectral confocal photoluminescence (PL) imaging is performed to fully characterize the atomically thin \WSe{} flake and the effects of strain perturbations. Figure~\ref{fig:monolayer-small}f shows a color-coded spatial map of the integrated intensity of the PL spectrum ($660-830 \unit{nm}$). The PL intensity significantly increases at the locations of the nanopillars. The highly reduced PL signal in between the nanopillar sites is likely due to efficient diffusion of excitons towards the lower energy states at the strain-tuned sites\cite{Feng12, Castellanos13local, Li15}. The spectra from several nanopillar sites as well as typical 1L and 2L regions between the nanopillars is exhibited in Fig.~\ref{fig:monolayer-small}g. A broad defect band with weak intensity is observed in the homogeneous 1L and 2L regions (at low excitation powers the 2D excitons are not visible). On the other hand, for each spectrum obtained at a location of a nanopillar (including nanopillar \#1 in the 2L region), a few discrete narrow linewidth ($< 200 \unit{\micro eV}$) peaks with high intensity are typically observed. Each of these peaks signify emission from single quantum emitters\cite{Srivastava15,He15,Koperski15,Chakraborty15,Tonndorf15,Kumar15,Kern16,Kumar16,Palacios16}. A second-order correlation measurement is presented in Fig.~\ref{fig:monolayer-small}e, where a fit to the data yields $\gt(0) = 0.07 \pm 0.04$ with a decay time of $\tau = 2.8 \pm 0.2 \unit{ns}$. This demonstrates photon antibunching and the quantum nature of the 1L discrete lines.

\begin{figure*}[t]
   	\centering
	\includegraphics[trim={1.4cm 19.5cm 1.4cm 1.4cm},clip, width=\textwidth]{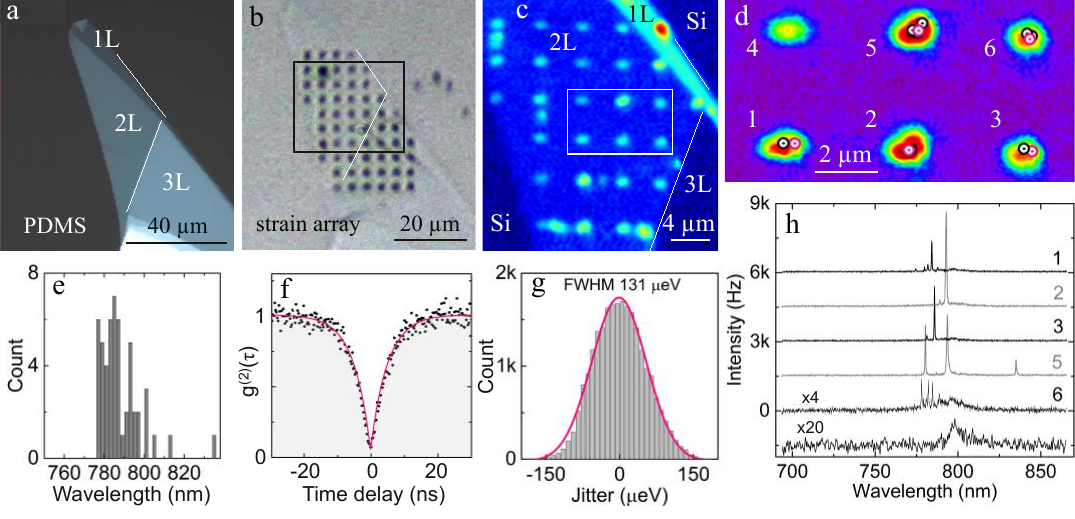}
   	\caption{Optical micrograph of 2L WSe2 (a) before and (b) after the transfer onto the nanopillars. (c) A 2D spatial map of the PL intensity within $700 - 860\unit{nm}$. (d) A high resolution spatial map of integrated intensity of the 6 nanopillars indicated in (c). The circles mark the positions of emitters in (h). (e) A histogram of the wavelengths of the 2L \WSe{} emitters. (f) Photon auto-correlation histogram from the bi-layer emitter at nanopillar \#1 of Fig. 1. A fit using $\gt(0) = 0.03 \pm 0.02$ and $\tau = 4.8 \pm 0.1 \unit{ns}$ is shown. (g) A histogram of the spectral jitter over 20 hours (3 s time-bin) of the emitter at nanopillar \#1 of Fig. 1. (h) PL spectra from the emitter positions identified in (d) with red circles.}
   	\label{fig:bilayer}
\end{figure*}

To demonstrate the universality of nanoscale strain engineering to generate strain-induced quantum emitters in atomically thin semiconductors, we create an array of pure single photon emitters in 2L \WSe{}. Bi-layer \WSe{} is an intriguing host for quantum emitters as it offers an additional pseudo-spin based on the layer degree of freedom\cite{Jones14}. While 2L \WSe{} is an intrinsically indirect band gap semiconductor, the indirect and direct transitions are nearly degenerate and under tensile strain the band structure can be modified such that the direct transition becomes preferred\cite{Desai14}. We exfoliate a flake with a large 2L region (see Fig.~\ref{fig:bilayer}a) and transfer it to an array of nanopillars (with h:w $\sim 0.3$) on a Si substrate (Fig.~\ref{fig:bilayer}b). Once again, we observe a huge increase in PL intensity at the nanopillar sites (Fig.~\ref{fig:bilayer}c), evidencing the transition to a direct electronic gap and the exciton funnel effect due to local strain. We note there is also a small 1L and a large 3L region of the flake in Fig.~\ref{fig:bilayer}b that cover nanopillar sites. While the 1L region shows similar properties to the flake in Fig.~\ref{fig:monolayer-small}a, the 3L remains dark in PL. Crucially, bright narrow-linewidth spectral lines (Fig.~\ref{fig:bilayer}h) are again observed at the strain-induced sites. Figure~\ref{fig:bilayer}d shows a high-resolution spatial map of 6 nanopillars in the center of the array superimposed with the locations of the quantum emitters in Fig.~\ref{fig:bilayer}h. A wavelength histogram ($2 \unit{nm}$ binning) of all strain-induced emitters (53 in total) created in the 2L \WSe{} array is shown in Fig.~\ref{fig:bilayer}e. The histogram shows that emitters span a wavelength region from $775\unit{nm}$ to $835\unit{nm}$. A Gaussian fit to the data is used to quantify the inhomogeneous distribution of the emitters, yielding $33 \unit{meV}$ full width at half maximum (FWHM). Notably, several emitters at very similar wavelengths are observed (e.g. in Fig.~\ref{fig:bilayer}h peaks at $793 \unit{nm}$ in spectra 2 and 5; peaks at $785 \unit{nm}$ in spectra 1 and 3).  The second order correlations from each individual peak measured exhibit highly pure single photon emission, e.g the single peak at $\lambda = 801.08 \unit{nm}$ from nanopillar \#1 in Fig.~\ref{fig:monolayer-small}g shows $\gt(0) = 0.03 \pm 0.02$ with a decay time of $\tau = 4.8 \pm 0.1 \unit{ns}$ (Fig.~\ref{fig:bilayer}f). For spectra recorded over a 20 hour period, this emitter does not blink. Figure~\ref{fig:bilayer}g shows a histogram of the spectral jitter recorded over 20 hours (using 3 s time-bin) from the 2L emitter at nanopillar \#1. The histogram is fit by a Gaussian distribution with $131 \unit{\micro eV}$ FWHM. Further, these quantum emitter arrays are optically stable and robust, surviving multiple sample cooling and heating cycles.

\begin{figure*}[t]
   	\centering
	\includegraphics[trim={1.4cm 18.7cm 1.4cm 1.4cm},clip, width=\textwidth]{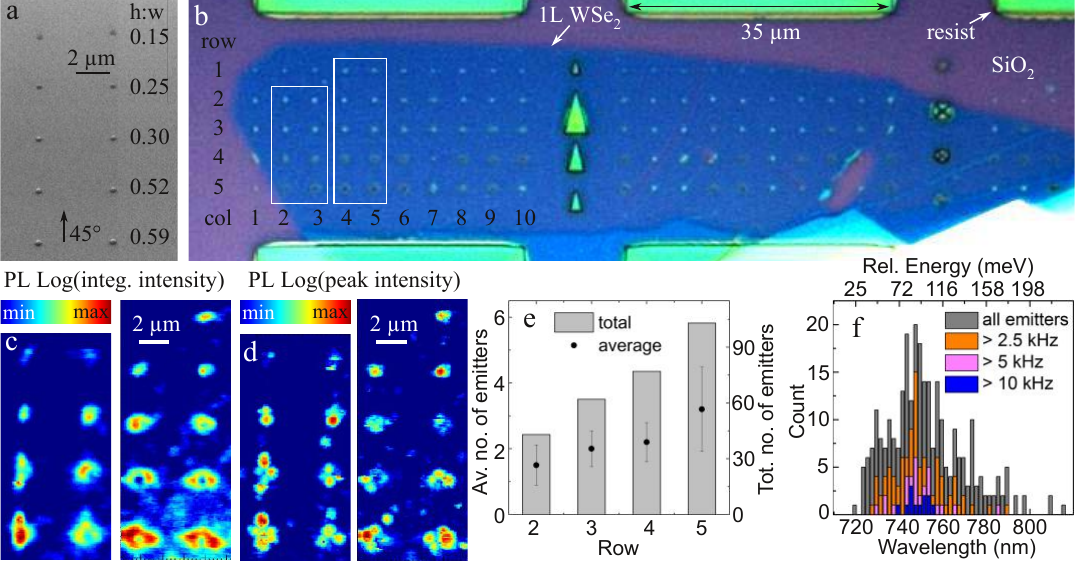}
   	\caption{(a) A $45\degree$ SEM image of a SiO$_2$ substrate with an array of nanopillars of varying aspect ratios, as labelled. (b) Optical micrograph of a large ($100 \times 25 \unit{mm}$) 1L \WSe{} flake covering 101 nanopillars. Some high aspect ratio nanopillars exhibit a dark center with a bright ring, which signifies the pillar punctured the flake during the transfer process. (c) High-resolution color-coded spatial map of integrated PL signal in the spectral range of $690 -850 \unit{nm}$ for columns 2 to 5 from the region outlined in (b). In the case of punctured flake at a nanopillar site there is a clear ring in the PL intensity. (d) Same maps as (c) showing the peak intensity. The individual emitters are resolved. (e) Statistics on the emitters per pillar for each row. (f) Histogram of the emitter wavelengths with post-selection on the emitter brightness (peak intensity).}
   	\label{fig:monolayer-big}
\end{figure*}

Having established a technique to successfully create robust strain-induced quantum emitters, we seek to optimize the process. We tailor the local elastic strain by varying the height to change the aspect ratio of five rows of nanopillars, from h:w 0.15 to 0.59, as shown in the scanning electron micrograph in Fig.~\ref{fig:monolayer-big}a. Precise measurements of the pillar size are made by AFM (see Supp. Info. Fig.~S3). Figure~\ref{fig:monolayer-big}b shows an optical micrograph of a large ($\sim 100 \times 25 \unit{\micro m}$) 1L \WSe{} flake covering 101 nanopillars in the array. Here the pillar locations are identified by the bright points in the micrograph. Also visible in Fig.~\ref{fig:monolayer-big}b in some cases for the high aspect ratio nanopillars (e.g. Row 5, Columns 3--5) is a small ring surrounding the pillar. We show in the following that in such cases the pillar has punctured the 2D flake during the transfer process. Flake puncturing is not observed for any of the low aspect ratio nanopillars in Rows 1 and 2 or with the nanopillars used in Figs.~\ref{fig:monolayer-small} and \ref{fig:bilayer}. Figure~\ref{fig:monolayer-big}c shows a spatial map of the integrated intensity (logarithmic scale) of the entire PL spectrum ($690-850 \unit{nm}$) from Columns 2 -- 5 (as labelled in Fig.~\ref{fig:monolayer-big}b). PL maps of the entire flake with several spectra are shown in the Supplementary Information (Fig.~S4). A mixture of peaks with large ($700-900 \unit{\micro eV}$) fine-structure splittings (e.g. Fig. S5) and single linearly polarised peaks are observed, as reported previously\cite{Kumar16}. The PL intensity at the nanopillars increases with increasing nanopillar aspect ratio, as expected for the diffusion of excitons towards the local regions with strain-tuned band-gaps. Further, in the cases where the flake is pierced, low intensity PL at the nanopillar center is surrounded by a ring of intense PL. 

Figure~\ref{fig:monolayer-big}d maps the locations of the intense, narrow linewidth peaks that signify quantum light emission in Columns 2 -- 5. We determine the positioning accuracy of the strain-induced quantum emitters by comparing their location to the center of the nanopillars (see Fig.~S6 for the measurement of the nanopillar centre). We observe that when the flake is pierced by a nanopillar, the emitters are found at the circumference of the nanopillar. On the other hand, the quantum emitters are created in the center of the nanopillars for flakes that conform to the substrate topography. In particular, the positioning accuracy of the emitters in Row 2 reveal high precision for positioning accuracy: $\sigma_D^\text{Row 2} = 120 \pm 32 \unit{nm}$, where $\sigma_D$ is the average displacement of the emitter from the nanopillar center.  Displacement statistics of the emitters in Rows 3 -- 5 are: $\sigma_D^\text{Row 3} = 262 \pm 46 \unit{nm}$, $\sigma_D^\text{Row 4} = 476 \pm 85 \unit{nm}$, and $\sigma_D^\text{Row 5} = 521 \pm 64 \unit{nm}$. Row 1 did not yield sufficient emitter numbers for useful statistics. In the best case (Column 5, Row 2), the displacement is $30 \unit{nm}$. While this accuracy is sufficient to couple to cavity or waveguide modes for cavity quantum electrodynamics experiments and integrated quantum photonics applications\cite{Vahala03,Lodahl15}, further improvements for more precise and robust positioning of single emitters is likely possible with smaller diameter nanopillars that maintain an aspect ratio similar to Row 2.  

Statistics of all emitters identified (285 in total) in the entire 1L \WSe{} flake over the 101 nanopillars is shown in Fig.~\ref{fig:monolayer-big}e and f. Figure~\ref{fig:monolayer-big}e shows the statistics for the number of emitters per pillar. While distinct emitters are sometimes found at the nanopillar sites in Row 1 (e.g. the spectra in Fig.~S4c), they tend to be difficult to distinguish from the broad background. With the increased aspect ratio of the nanopillars in Row 2, one or two distinct bright emitters are found with a yield of 85\% (17 of 20 nanopillars contain pure single emitters). Rows 3 and 4 have near unity yields: 96\% (45 of 47 nanopillars yield at least 1 quantum emitter). Figure~\ref{fig:monolayer-big}f shows the emitter wavelength histogram. While the overall energy distribution of the single emitters is broad (spanning $\sim 200 \unit{meV}$), all emitters emit at lower energy than the bright 2D exciton peak in \WSe{} ($\sim 1.74 \unit{eV}$\cite{Wang14}). The emitter wavelength histogram is also post-selected for emitter brightness (the peak intensity at $\sim 0.8$ saturation power) for this flake. Fitting a Gaussian distribution to the histogram gives $73 \unit{meV}$ FWHM, more than twice as large as the 2L emitters.

Layered transition metal dichalcogenide semiconductors are attractive hosts for quantum emitters due to the unique valley degree of freedom and strong spin-orbit coupling. Using nanoscale elastic strain engineering, we have achieved local modification of the electronic and optical properties to deterministically generate robust quantum emitters in this emerging quantum photonic platform. The straightforward fabrication procedure presented here is scalable and likely to be applicable to other 2D materials. Notably, we observe negligible background signal at the base of the discrete spectral lines, enabling high purity single photon emission in both 1L and 2L \WSe{}. Compared to mono-layer \WSe{}, the bi-layer appears additionally attractive as a host for single excitons due to a narrower inhomogenous distribution and the potential for spin-layer locking.

\bigskip

We thank N. Ross for assistance with electron beam lithography. This work was supported by a Royal Society University Research Fellowship, the EPSRC (grant numbers EP/I023186/1, EP/K015338/1, EP/L015110/1, and EP/M013472/1) and an ERC Starting Grant (number 307392).

%


\clearpage
\onecolumngrid

\setcounter{figure}{0}

\renewcommand{\thefigure}{S\arabic{figure}}

\begin{center}
\large\textbf{\textit{Supplementary material:} Deterministic strain-induced arrays of quantum emitters \\
in a two-dimensional semiconductor}
\end{center}
 
This supplementary information gives insight into detailed aspects of the results. These additional measurements allow for better understanding of the samples and quantum emitters behaviour in monolayer and bilayer \WSe{}. 

\begin{figure*}[h!]
   	\centering
	\includegraphics[trim={1.4cm 21cm 1.4cm 1.4cm},clip, width=\textwidth]{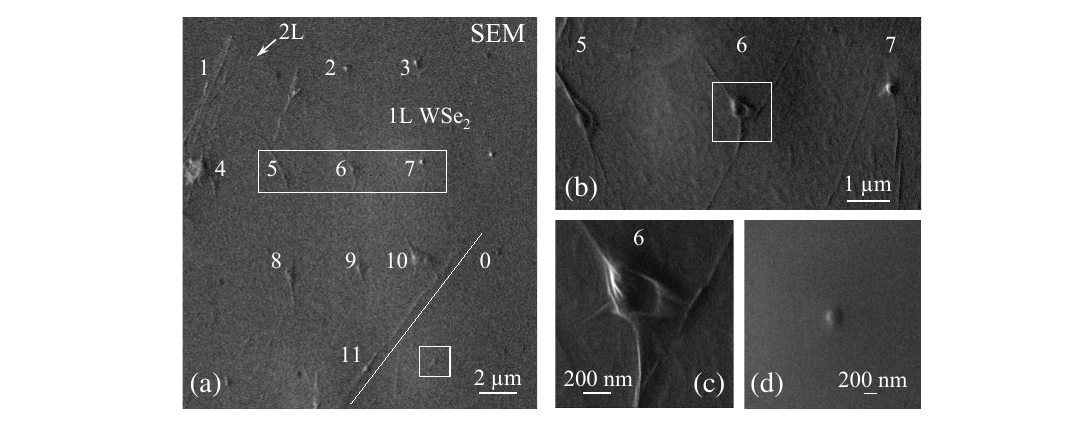}
   	\caption{SEM images of the monolayer \WSe{} flake on nanopillars presented in Fig. 1 of main article. (a) The entire flake with the region of interest. (b) Magnification of nanopillars 5, 6 and 7. (c) Magnification of nanopillar 6. The wrinkling of the monolayer at the nanopillar correlates with that measured by AFM in the main manuscript. (d) A bare nanopillar is shown for comparison. It is identified at the bottom right of (a) by a white square.}
   	\label{fig:monolayer-small-SEM}
\end{figure*}

Figure \ref{fig:monolayer-small-SEM} shows scanning electron microscope (SEM) images of a monolayer \WSe{} transferred onto the nanopillars. The topography observed here matches that observed in the atomic force microscope (AFM) image of Fig.~1c in the main article. A magnification of the areas framed in Fig.~\ref{fig:monolayer-small-SEM}a is presented in Fig.~\ref{fig:monolayer-small-SEM}b, c and d. The wrinkled monolayer around nanopillar 6 is clearly visible in Fig.~\ref{fig:monolayer-small-SEM}c. Figure \ref{fig:monolayer-small-SEM}d shows a bare nanopillar for comparison (identified by a white square at the bottom right of Fig.~\ref{fig:monolayer-small-SEM}).

\begin{figure*}[h!]
   	\centering
	\includegraphics[trim={1.4cm 24cm 1.4cm 1.4cm},clip, width=\textwidth]{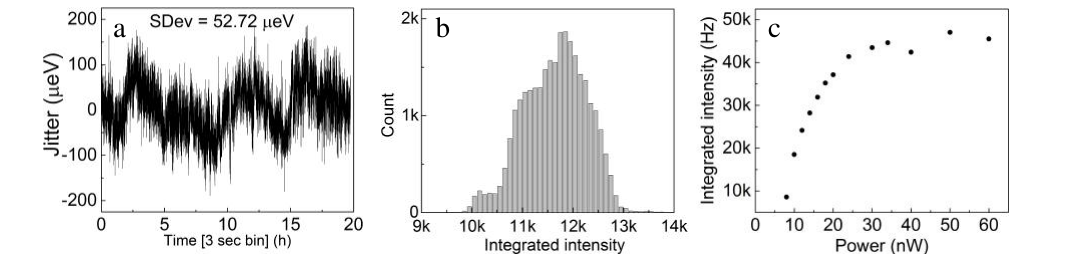}
   	\caption{(a) Time-trace of the peak energy detuning of the photoluminescence of emitter 1 in Fig.~1 in a bilayer flake of \WSe{}, using $3 \unit{s}$ time bins. (b) Histogram of intensity performed on the previous time-trace. (c) Power dependence of the integrated intensity emitted by this emitter, showing a typical saturation behaviour.}
   	\label{fig:bilayer-emitter-in-depth}
\end{figure*}

Figure \ref{fig:bilayer-emitter-in-depth} characterizes more precisely the 2L emitter 1 of Fig.~1 of the main article. Figure~\ref{fig:bilayer-emitter-in-depth}a shows a time-trace of the peak energy over 20 hours of measurement. The standard deviation is $53\unit{\micro eV}$), showing good emission energy stability. Figure~\ref{fig:bilayer-emitter-in-depth}b is an intensity histogram performed on the same time-trace measurement. It accounts for an excellent intensity stability with no blinking observed. The majority of intensity fluctuations are due to vibrations during the period. Finally, Figure~\ref{fig:bilayer-emitter-in-depth} is a power dependence of the integrated photoluminescence intensity. The emitter exhibits a very clear saturation behaviour consistent with a two-level system.

\clearpage

\begin{figure*}[h!]
   	\centering
	\includegraphics[trim={1.4cm 15cm 1.4cm 1.4cm},clip, width=\textwidth]{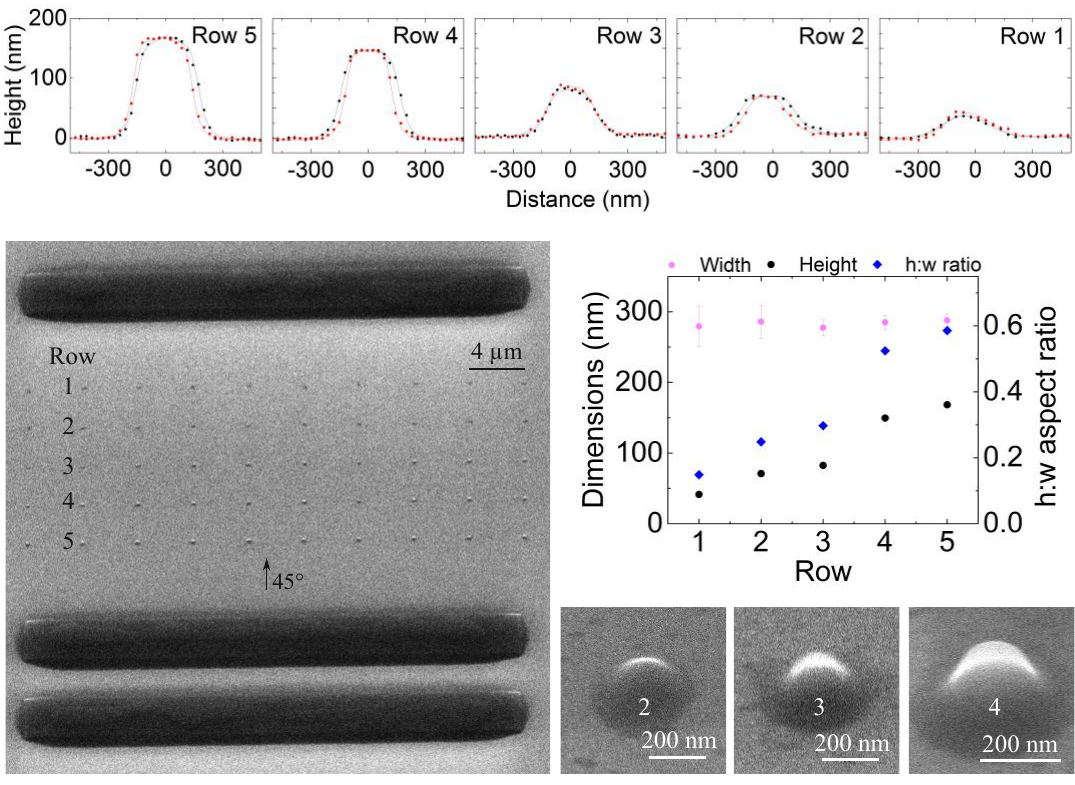}
   	\caption{(a) AFM profile of nanopillars for each row of the sample shown in Fig. 3 of the main article and \ref{fig:monolayer-big-spectra}. (b) Dimensions and aspect ratio of the nanopillars for each row, calculated from (a). (c) SEM image of the sample showing nanopillars of different heights for each row. (d) Detailed SEM image of nanopillars in rows 2, 3 and 4, respectively with aspect ratios 0.25, 0.30 and 0.52.}
   	\label{fig:monolayer-big-AFM-SEM}
\end{figure*}

\begin{figure*}[h!]
   	\centering
	\includegraphics[trim={1.4cm 10.cm 1.4cm 1.4cm},clip, width=\textwidth]{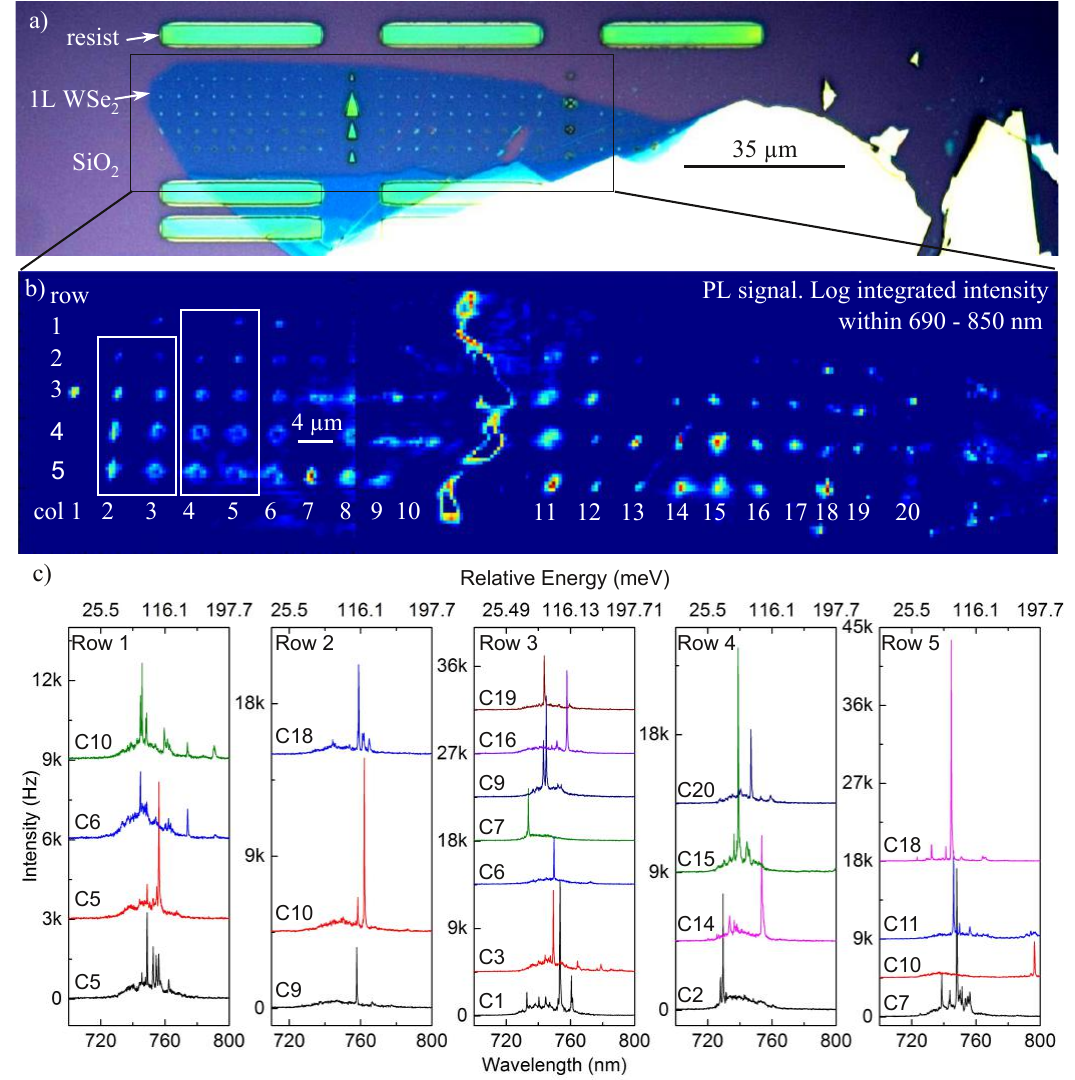}
   	\caption{(a) Optical micrograph of the flake of \WSe{} shown in Fig. 3 of the main article. (b) Spatial map of integrated PL signal showing the complete flake. The full map has 4 separate maps stitched together. During the second map, some motor hysteresis is visible. (c) PL spectra of individual emitters. The nanopillar position is indicated by a row-column reference (C = column).}
   	\label{fig:monolayer-big-spectra}
\end{figure*}

Figure~\ref{fig:monolayer-big-AFM-SEM} shows the nanopillars characterization on the same sample as Fig.~\ref{fig:monolayer-big-spectra}, prior to the deposition of the monolayer \WSe{}. Figure~\ref{fig:monolayer-big-AFM-SEM}a are vertical profiles of the nanopillars for each row. Figure~\ref{fig:monolayer-big-AFM-SEM}b displays the dimensions and aspect ratio extracted from Fig.~\ref{fig:monolayer-big-AFM-SEM}a. The width stays constant at $280\pm 20 \unit{nm}$ for all rows while the height increases from $40\unit{nm}$ to $170\unit{nm}$. Figure~\ref{fig:monolayer-big-AFM-SEM}c is an SEM image of the sample, where the different rows of nanopillars are visible corresponding to different aspect ratios. Figure~\ref{fig:monolayer-big-AFM-SEM}d is a magnification of nanopillars from rows 2, 3 and 4. The change in height is clearly visible.

\clearpage

Figure~\ref{fig:monolayer-big-spectra} shows more extensive data on the monolayer \WSe{} sample displayed in Fig.~3 of the main article. Figure~\ref{fig:monolayer-big-spectra}a shows the larger view of the same optical micrograph as Fig.~3b. Figure~\ref{fig:monolayer-big-spectra}b shows 4 different spatial maps stitched together displaying the integrated photoluminescence intensity on the color scale. PL intensity is systematically enhanced at the position of the nanopillars. Figure~\ref{fig:monolayer-big-spectra}c shows several photoluminescence spectra for diverse quantum emitters on the position of the nanopillars.

\clearpage

\begin{figure*}[h!]
   	\centering
	\includegraphics[trim={1.4cm 21.cm 1.4cm 1.4cm},clip, width=\textwidth]{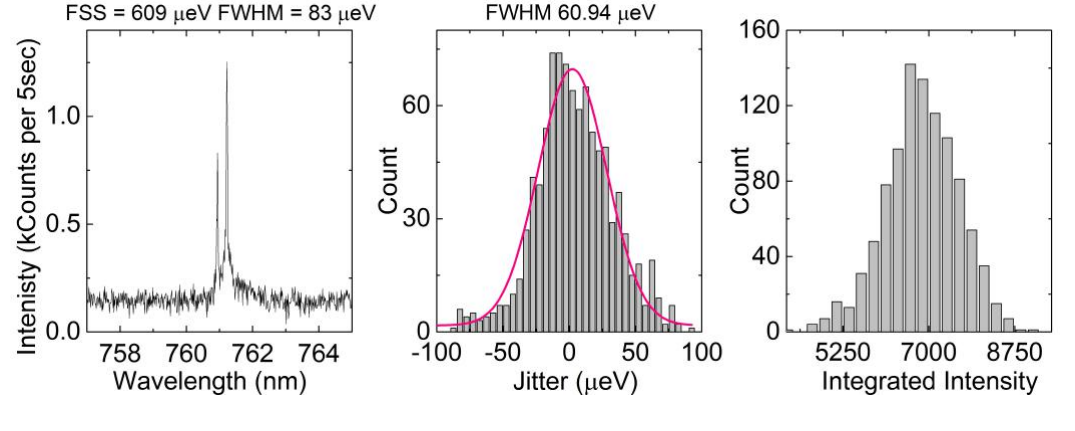}
   	\caption{Left: Photoluminescence spectrum showing fine-structure splitting, extracted from a time-trace measurement. Center: Histogram of detuning for the time-trace measurement ($82\unit{min}$ measurement with $5\unit{s}$ time bins). Right: Histogram of the integrated intensity showing PL stability during the time-trace measurement.}
   	\label{fig:fss}
\end{figure*}

Figure \ref{fig:fss} shows the results from a time-trace measurement from the emitter at nanopillar \#2 in the 1L flake from Fig.~1 in the main text. The left panel is a PL emission spectrum presenting a fine structure-splitting of $600\unit{\micro eV}$. This spectrum is part of a time-trace measurement of $82\unit{min}$ with $5\unit{s}$ time bins. The center graph is an energy histogram integrated from the time-trace. The PL peak energy fluctuates $60\unit{\micro eV}$. Finally, the right panel is an intensity histogram showing the non-blinking behaviour, with a mean intensity of $7000$ counts per time bin.

\clearpage

The position of the nanopillars in the hyperspectral spatial maps is not known a priori. We therefore use a spectral weighted averaging (WA) method to determine the centre position of the nanopillars. 

\begin{figure*}[h!]
   	\centering
	\includegraphics[trim={1.4cm 20.cm 1.4cm 1.3cm},clip, width=\textwidth]{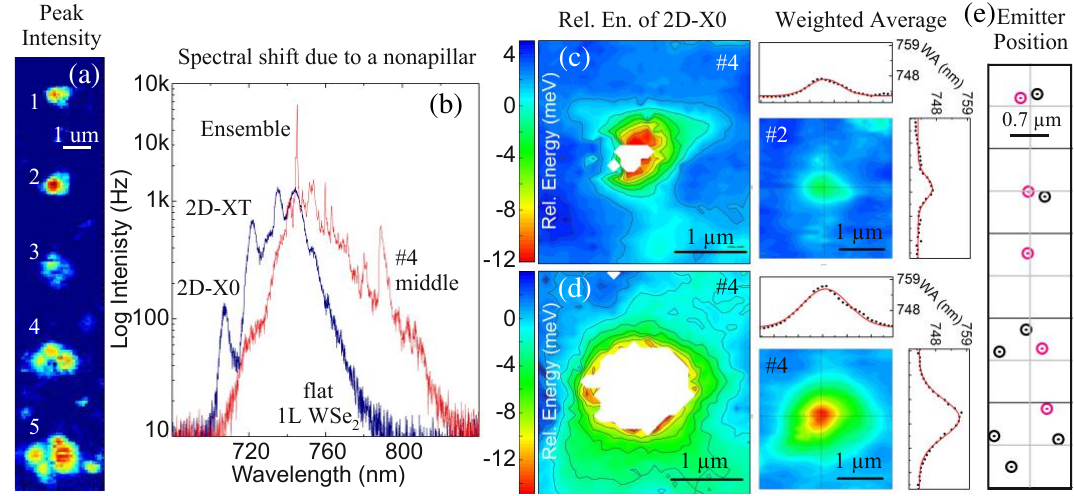}
   	\caption{Weight averaging technique used for determining the centre of the nanopillars. (a) High resolution spatial map showing the peak intensity in the PL spectra from Column 5 of the \WSe{} on SiO$_2$ sample shown in Fig.~\ref{fig:monolayer-big-spectra} and Fig.~3 of the main article. The individual emitters are resolved. (b) Two typical PL spectra, one from the nanopillar and one off the nanopillar in an unstrained region. The color-coded spatial maps of of the relative energy of 2D-X0 peak and of the weighted average for pillars \#2 (c) and \#4 (d)from column 5. The left map displays the shift in 2D exciton energy with the respect to the signal measured from flake on flat (off the nanopillar). The weighted average map probes the spectral shift in energy by taking into account the entire spectrum. The position of the nanopillars were determined by fitting Gaussian to the WA profile, shown by the cross-sections. (e) Positions (circle markers) of the single emitters from (a) relative to the centre of the nanopillars (grey crosses).}
   	\label{fig:wa}
\end{figure*}

Figure \ref{fig:wa} illustrates this process. First, Fig.~\ref{fig:wa}a shows a high definition spatial map of the spectral peak intensity. The individual emitters are resolved but the accurate position of the centre of the nanopillars is unknown. Figure~\ref{fig:wa}b shows two PL spectra: one off the nanopillar and one on a nanopillar. Strain induces a global shift in the energy of the spectrum. We assume that the strain experienced by the flake is at its highest in the middle of the nanopillar. Thus, taking the mean emission energy and mapping its shift with position enables us to determine the position of the nanopillars with high accuracy. Figures~\ref{fig:wa}c and d show spatial maps for the energy of the 2D exciton (2D-X0) on the left and average wavelength weighted by the intensity on the right. The weighted average $WA$ is computed using:
\begin{equation}
WA = \frac{\sum_\lambda\lambda\cdot I}{\sum_\lambda{I}}
\end{equation}
where $I$ is the intensity.

The energy of the 2D exciton also shifts with strain, and has been previously used to estimate the strain\cite{Kumar15}. However, only the $WA$ can fully map the nanopillar in the data presented here. The white region in the middle of 2D-X0 is where the 2D-X0 intensity vanishes at higher strains. On the other hand, the profile of weighted average yields a smooth, clean contour map from which the centre of nanopillars can be accurately determined. 

Figure~\ref{fig:wa}e shows the assessed positions of the emitters (circle markers) compared to the position of the centre of the nanopillars in Column 5 (grey crosses) determined using the weighted average method.

\end{document}